\documentclass[pdflatex,sn-nature]{sn-jnl}

\usepackage{graphicx}
\usepackage{multirow}
\usepackage{amsmath,amssymb,amsfonts}
\usepackage{amsthm}
\usepackage{mathrsfs}
\usepackage[title]{appendix}
\usepackage[usenames,dvipsnames]{xcolor}
\usepackage{textcomp}
\usepackage{manyfoot}%
\usepackage{booktabs}%
\usepackage{algorithm}%
\usepackage{algorithmicx}%
\usepackage{algpseudocode}%
\usepackage{listings}%

\usepackage{latexsym}
\usepackage{subcaption}
\usepackage{balance}
\usepackage{caption}
\usepackage{esdiff}
\usepackage{ragged2e}
\usepackage{lineno}
\usepackage{gensymb}
\usepackage{enumerate}
\usepackage{hyperref}
\usepackage{graphics}

\usepackage{float}
\renewcommand{\thefigure}{\arabic{figure}}
\renewcommand{\thetable}{\Roman{table}}
\renewcommand{\theequation}{\arabic{equation}}
\newcommand{\bea}{\begin{eqnarray}}
\newcommand{\eea}{\end{eqnarray}}
\newcommand{\cm}{~cm$^{-1}$}

\begin{document}

\title{\textbf{Magneto-Optical Study of Chiral Magnetic Modes in NiI$_{2}$: Direct Evidence for Kitaev Interactions}}

\author[1]{\fnm{Kartik} \sur{Panda}}
\equalcont{These authors contributed equally to this work.}

\author[2]{\fnm{Chaebin} \sur{Kim}}
\equalcont{These authors contributed equally to this work.}

\author[1]{\fnm{Daniel} \sur{Bazyliansky}}
\equalcont{These authors contributed equally to this work.}

\author[3]{\fnm{Javier} \sur{Taboada-Gutiérrez}}

\author[4]{\fnm{Florian} \sur{Le Mardelé}}

\author[4,5]{\fnm{Jan} \sur{Dzian}}

\author[1]{\fnm{Guy} \sur{Levy}}

\author[6]{\fnm{Jae Ha} \sur{Kim}}

\author[7]{\fnm{Youjin} \sur{Lee}}

\author[8]{\fnm{Bumchan} \sur{Park}}

\author[2]{\fnm{Martin} \sur{Mourigal}}

\author[8]{\fnm{Jae Hoon} \sur{Kim}}

\author[3]{\fnm{Alexey} \sur{B. Kuzmenko}}

\author[4]{\fnm{Milan} \sur{Orlita}}

\author*[7]{\fnm{Je-Geun} \sur{Park}}\email{jgpark10@snu.ac.kr}

\author*[1]{\fnm{Nimrod} \sur{Bachar}}\email{nimib@ariel.ac.il}

\affil[1]{\orgdiv{Department of Physics}, \orgname{Ariel University}, \orgaddress{\city{Ariel}, \postcode{4070000}, \country{Israel}}}

\affil[2]{\orgdiv{School of Physics}, \orgname{Georgia Institute of Technology}, \orgaddress{\city{Atlanta}, \postcode{30332}, \state{Georgia} \country{USA}}}

\affil[3]{\orgdiv{Department of Quantum Matter Physics}, \orgname{Université de Genève}, \orgaddress{\street{24 Quai Ernest Ansermet}, \city{Geneva}, \postcode{CH-1211}, \country{Switzerland}}}

\affil[4]{\orgdiv{LNCMI-EMFL, CNRS UPR3228}, \orgname{Université Grenoble Alpes, Université Toulouse, Université Toulouse 3, INSA-T}, \orgaddress{\city{Grenoble and Toulouse}, \country{France}}}

\affil[5]{\orgdiv{Institute of Physics}, \orgname{Charles University}, \orgaddress{\street{Ke Karlovu 5}, \city{Prague}, \postcode{121 16}, \country{Czech Republic}}}

\affil[6]{\orgdiv{Center for Nano Materials}, \orgname{Sogang University}, \orgaddress{\city{Seoul}, \postcode{04107}, \country{Republic of Korea}}}

\affil[7]{\orgdiv{Department of Physics and Astronomy}, \orgname{Seoul National University}, \orgaddress{\city{Seoul}, \postcode{08826}, \country{Republic of Korea}}}

\affil[8]{\orgdiv{Department of Physics}, \orgname{Yonsei University}, \orgaddress{\city{Seoul}, \postcode{03722}, \country{Republic of Korea}}}


\abstract{
\textbf{
Bond-dependent magnetic interactions, particularly those described by the Kitaev model, have emerged as a key pathway toward realizing unconventional magnetic states such as quantum spin liquids and topologically nontrivial excitations, including skyrmions. These interactions frustrate conventional magnetic order and give rise to rich collective behavior that continues to challenge both theory and experiment. While Kitaev physics has been extensively explored in the context of honeycomb magnets, direct evidence for its role in real materials remains scarce. Magnetic van der Waals (vdW) materials have emerged as a versatile platform for exploring low-dimensional electrical, magnetic, and correlated electronic phenomena~\cite{Park2016,Burch2018,Gibertini2019}, and provide a fertile ground for potential applications ranging from spintronics to multiferroic devices and quantum information technologies~\cite{Novoselov2016}. Here, we demonstrate, through magneto-transmission, Faraday angle rotation, and magnetic circular dichroism measurements, that the magnetic excitation spectrum of NiI$_2$, a van der Waals multiferroic material, is more accurately captured by a Kitaev-based spin model than by the previously invoked helical spin framework.}}


\maketitle





NiI$_2$ has a layered structure comprising edge-sharing [NiI$_6$] octahedra separated by van der Waals (vdW) gaps~\cite{Son2022,Botana2019,Pasternak1990,Kapeghian2024}. It hosts a non-centrosymmetric, helically ordered antiferromagnetic state below two transition temperatures, $T_{\mathrm{N1}} = 76$~K and $T_{\mathrm{N2}} = 59$~K~\cite{Pasternak1990,Kurumaji2013,Billerey1977,Billerey1980,Kuindersma1981}. The stability of magnetism in NiI$_2$ reflects the presence of anisotropies and spin–orbit interactions that lift degeneracy and enable long-range order even in quasi-two-dimensional systems~\cite{Huang2017,Gong2017,Song2022}, in contradiction to the Mermin–Wagner theorem~\cite{Mermin1966}. In a broader context, recent interest in altermagnets, materials with non-relativistic spin polarization and broken time-reversal symmetry, has emerged as a key area of research, particularly for their potential in spintronic and quantum information applications. NiI$_2$, recently classified as a non-collinear $p$-wave magnet, provides new opportunities to explore symmetry-protected electrical control of spin polarization~\cite{Song2025}.

Two low-energy electromagnon modes at 34 and 37~cm$^{-1}$ have been observed through Raman spectroscopy~\cite{Song2022} and THz magneto-transmission measurements~\cite{Kim2023} in the helical antiferromagnetic (AFM) state. Raman spectroscopy shows that these modes exhibit opposite circular conductivity at zero magnetic field, while second harmonic generation (SHG) measurements suggest the breaking of both inversion and rotational symmetries in the helical AFM state~\cite{Wu2023}. Magneto-transmission measurements reveal that these chiral excitations redshift with increasing temperature~\cite{Kim2023,Song2022}, acting as soft modes of the magnetic transition, but exhibit a pronounced blueshift in response to an external magnetic field~\cite{Kim2023}, consistent with their magnetic origin.

In typical antiferromagnetic resonance, such as that observed for magnons, the application of a magnetic field leads to a splitting into two modes with opposite circular polarizations~\cite{Keffer1952}, as seen in several van der Waals AFMs such as FePS$_3$~\cite{McCreary2020,Wyzula2022}. However, in the case of NiI$_2$, two distinct magnon modes are already present at zero magnetic field and only blueshift under the application of an external magnetic field. Previous studies have interpreted this behavior using a spin-helical magnon model~\cite{Song2022,Gao2024}, similar to what is observed in spiral magnets, to calculate the energies and dispersion relations of these split modes. Yet, spiral magnons typically exhibit strong directional anisotropy in their magnetic field dependence, as seen in the archetypal case of linarite~\cite{Rule2017,Gotovko2019}.

\begin{figure*}
    \centering
    \includegraphics[width=1.0\textwidth]{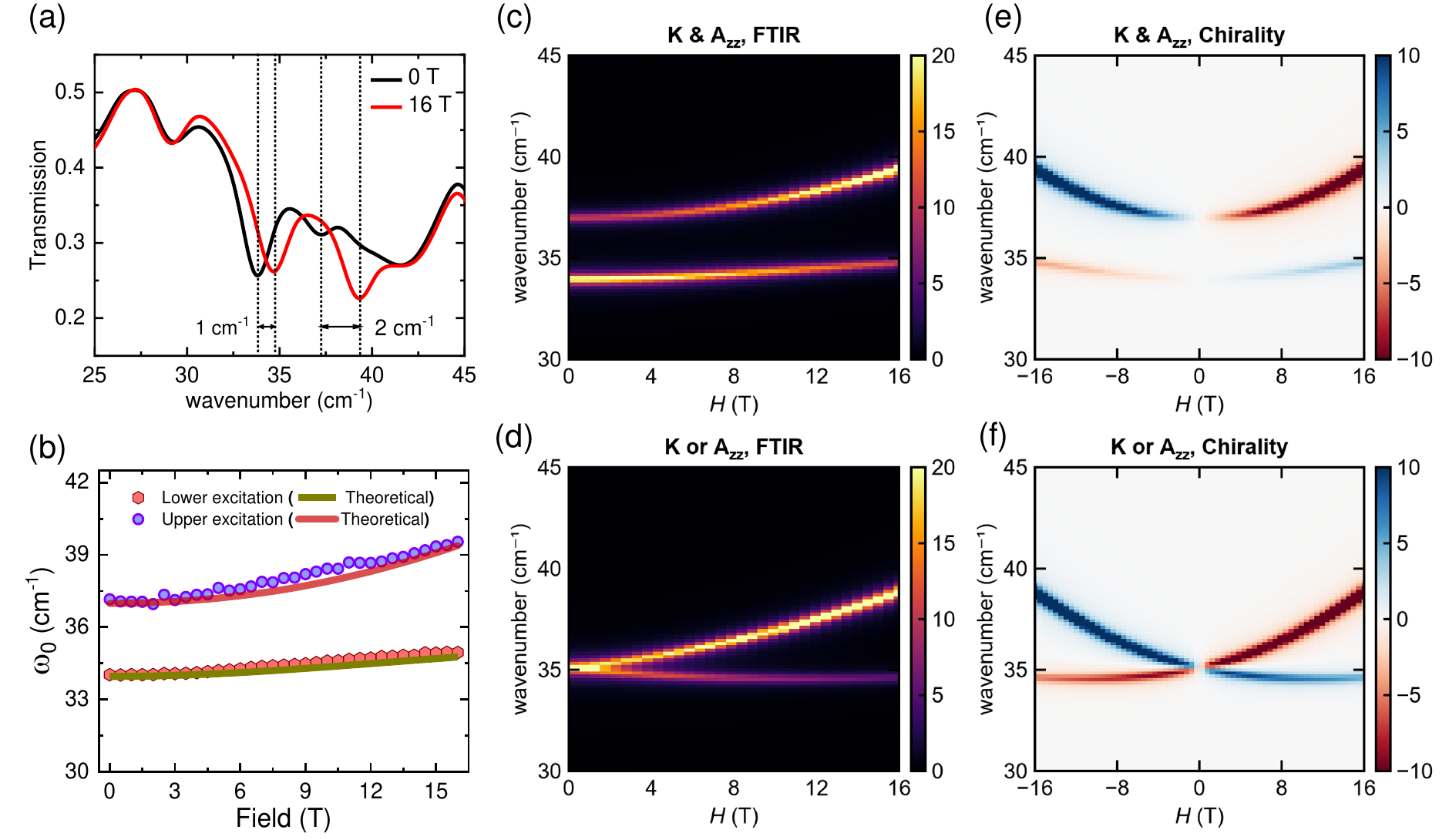}
    \captionsetup{justification=justified, singlelinecheck=false}
    \caption{
        \textbf{Magneto-transmission and chiral dynamical simulations of NiI$_2$.}  
        \textbf{a,} Magneto-optical transmission in the far-infrared spectral range measured at $T = 4.2$~K for zero and maximum applied magnetic field $B$.  
        \textbf{b,} Evolution of the magnon mode with magnetic field. The solid lines show theoretical calculations incorporating anisotropy, as described in the text.  
        \textbf{c,e,} FTIR intensity and chiral dynamical spin structure factor simulations of NiI$_2$ using linear spin-wave theory (LSWT).  
        \textbf{d,f,} Simulations of NiI$_2$ using only Kitaev interaction ($K$) or easy-plane single-ion anisotropy ($A_{zz}$).
    }
    \label{Figure1}
\end{figure*}

These observations raise a central question: do these modes represent distinct chiral magnons, or are they the result of a single magnon split due to symmetry-breaking effects? To address this question, we performed magneto-optical transmission measurements at 4.2~K, deep in the helical AFM state. In contrast to linarite, the magneto-transmission spectra exhibit no angular dependence (see Appendix~\ref{sec:appendixD}), indicating that NiI$_2$ is optically isotropic in the helical antiferromagnetic state, and suggesting a different microscopic origin for the magnons. Figure~\ref{Figure1}(a) shows the transmission spectra of the sample measured at two different magnetic fields: 0~T (black curve) and 16~T (red curve) in the Faraday configuration (field perpendicular to the sample surface and parallel to the propagation of light). The spectra reveal two distinct absorption modes at approximately 34~cm$^{-1}$ and 37~cm$^{-1}$ at zero magnetic field, corresponding to the well-known electromagnons of NiI$_2$. Upon application of a magnetic field, these modes exhibit a blueshift, but the magnitude differs for each magnon, which could not have been resolved in previous low-field measurements~\cite{Kim2023}. The mode at 34~cm$^{-1}$ shifts by 1~cm$^{-1}$, and the mode at 37~cm$^{-1}$ shifts by about 2~cm$^{-1}$ up to a field of 16~T.

Figure~\ref{Figure1}(b) presents the evolution of the central frequency, $\omega_0$, of the electromagnon modes as a function of applied magnetic field up to $B = 16$~T. The field evolution of the magnons does not follow the easy-axis model, in which $[\omega_0^2(B) - \omega_0^2(0)] \propto B^2$ (See Appendix~\ref{sec:appendixA}). Together with the angle-independent spectra of the magnons, as shown in Appendix~\ref{sec:appendixC} this discrepancy suggests that an alternative microscopic model is needed to capture the full behavior.

To this end, we modeled the NiI$_2$ system using a spin Hamiltonian that incorporates a dominant Kitaev exchange interaction and an in-plane anisotropy term, $A_{zz}$, corresponding to two distinct crystal anisotropies in the NiI$_2$ structure. We performed linear spin-wave theory (LSWT) simulations using a recently proposed parameter set~\cite{kim2024nii2} to calculate the magnetic-field dependence of the magnon modes (see Methods). Our calculations show that the Kitaev or anisotropy term alone cannot reproduce the observed zero-field splitting (see Fig.~\ref{Figure1}(d)) from a single magnon branch. Either the Kitaev interaction or the easy-plane anisotropy can independently determine the rotation plane of the helical magnetic structure. However, when both terms coexist, their competition lifts the degeneracy and splits the original mode into two distinct rotating modes: one favoring the $x$-direction and the other the $\kappa$-direction, which lies between the $y$- and $z$-directions~\cite{Gao2024}. As a result, the combination of both terms not only explains the splitting of a single mode into two magnon excitations (see Fig.~\ref{Figure1}(c)), but also accurately reproduces the measured field dependence (see Fig.~\ref{Figure1}(b) for comparison with our experimental data). The dichotomy in the magnons' field dependence can be explained by an easy-plane model~\cite{Rezende2019}, which in this case is effectively captured by the in-plane anisotropy term $A_{zz}$. A previous study~\cite{kim2024nii2} did not fix the value of $A_{zz}$ in the paramagnetic phase. By fitting our magneto-transmission data in the helical antiferromagnetic state, we determine $A_{zz} = 0.32$~meV.

Using the same model, we also simulated the chiral character of the magnons through the chiral dynamical spin structure factor. Our calculations show that the two magnon modes exhibit opposite chirality at finite magnetic field in both cases (see Fig.~\ref{Figure1}(e,f)). This opposite chirality is typically observed in spiral magnetic systems. Under a finite magnetic field, time-reversal symmetry is broken, and the magnon energies become nonreciprocal ($E(k) \neq E(-k)$). In incommensurate magnetic structures, the spin-wave spectrum consists of three eigenmodes: $E_0(k)$, $E_{Q_m} = E_0(k + Q_m)$, and $E_{-Q_m} = E_0(k - Q_m)$, where $k$ is the momentum and $Q_m$ is the magnetic propagation vector. Due to the broken symmetry, $E_{Q_m}(k)$ and $E_{-Q_m}(k)$ acquire different energies and opposite chirality at $k = 0$. Consequently, the two magnon modes at the $\Gamma$ point, originating from $E_{\pm Q_m}$, exhibit opposite chirality and evolve differently under an applied magnetic field (see Fig.~\ref{Figure1}(c–f)).

\begin{figure*}
    \centering 
    \includegraphics[width=1.0\textwidth]{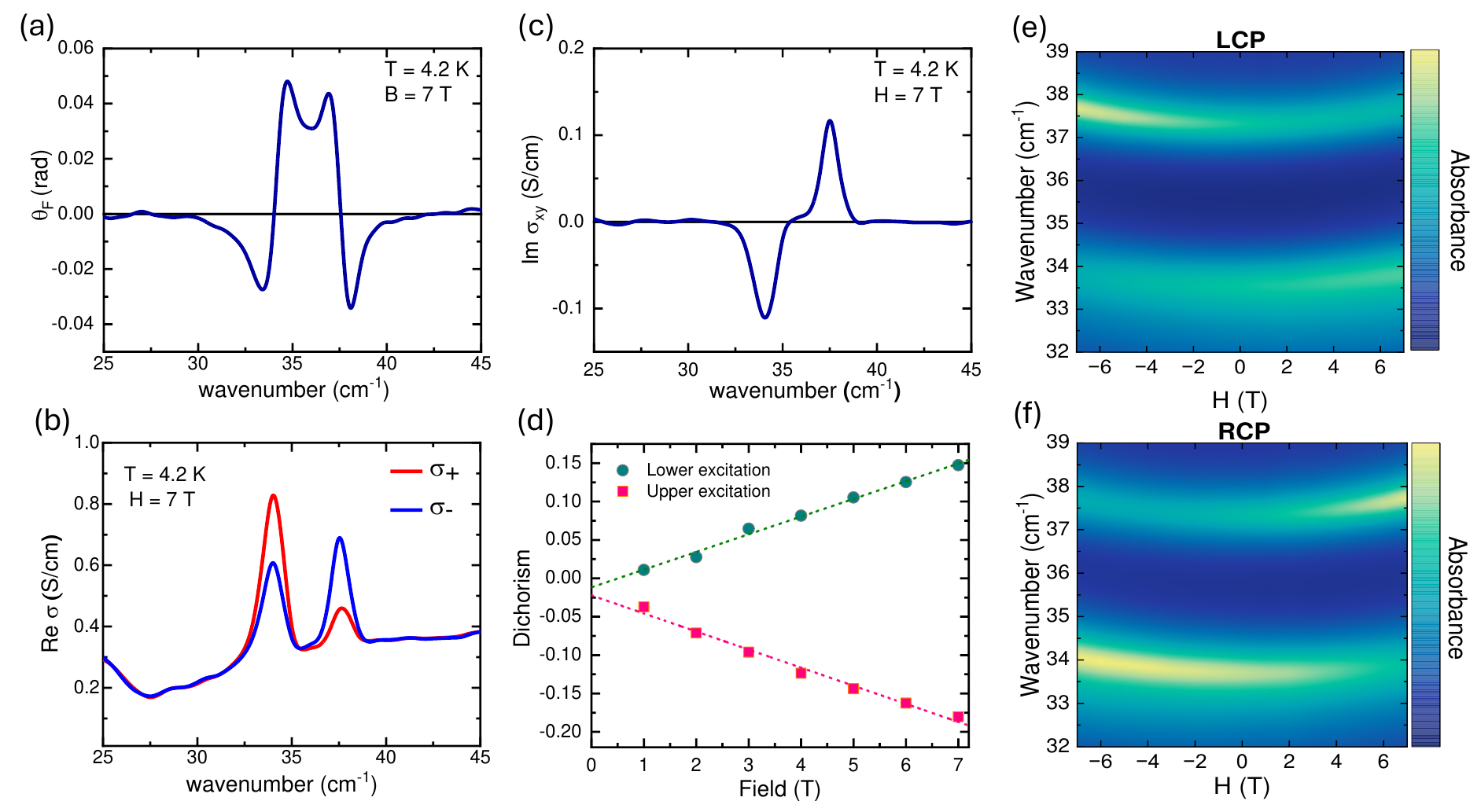}
    \captionsetup{justification=justified, singlelinecheck=false}
    \caption{
        \textbf{Faraday rotation and dichroism in NiI$_2$.}  
        \textbf{a,} Faraday rotation angle as a function of frequency at $T = 4.2$~K and an applied magnetic field of $B = 7$~T, estimated using the fast protocol described in the text.  
        \textbf{b,} Left-handed ($\sigma_{-}$) and right-handed ($\sigma_{+}$) optical conductivities derived from the Faraday rotation data in the FIR range at $T = 4.2$~K and $B = 7$~T.  
        \textbf{c,} Imaginary part of the transverse optical Hall conductivity, $\text{Im}[\sigma_{xy}]$, in the FIR range at $T = 4.2$~K and $B = 7$~T.  
        \textbf{d,} Dichroism of right-circularly polarized light ($\sigma_{+}$) and left-circularly polarized light ($\sigma_{-}$). Solid lines indicate linear fits to the data.  
        \textbf{e,} Absorption spectra of NiI$_2$ measured with left-circularly polarized (LCP) terahertz radiation at 1.5~K, showing the magnetic field dependence of the 34~cm$^{-1}$ and 37~cm$^{-1}$ electromagnon modes.  
        \textbf{f,} Corresponding spectra measured under right-circularly polarized (RCP) radiation.
    }
    \label{Figure2}
\end{figure*}

To probe this inverse chirality behavior, we conducted Faraday angle rotation and magnetic circular dichroism (MCD) measurements. The Faraday rotation measurements, reported here for the first time on NiI$_2$ electromagnons, allow us to extract the real and imaginary parts of the circularly polarized optical conductivities (see Methods). Figure~\ref{Figure2}(a) presents the Faraday rotation angle of NiI$_2$ measured at 4.2~K and magnetic fields of $B = \pm 7$~T. Several striking features are evident. The zero crossing of $\theta_F$ marks the central frequencies of the magnon modes. We observe a substantial Faraday rotation of approximately 80~mrad at 7~T and 4.2~K, well above the system's sensitivity limit of 2~mrad and clearly above the background, as shown in Fig.~\ref{Figure2}(a) (black line). Notably, the spectral dependence of the Faraday rotation is antisymmetric between the two magnons, indicating opposite circular optical conductivities for the modes.

To investigate the characteristics of the magnetic excitations, we combined the absolute transmission spectra and the Faraday rotation spectra to extract the optical circular conductivity of the electromagnons (see Methods). Figure~\ref{Figure2}(b) shows the real part of the optical conductivity, $\sigma_1$, separated into right- and left-handed circular polarization components, denoted $\sigma_{\pm}$. A clear dichroic response in the spectral weight of the circular optical conductivity is observed only near the electromagnon excitations. The lower mode exhibits approximately 40\% more spectral weight in the right-handed ($\sigma_{+}$, red) circular conductivity than in the left-handed ($\sigma_{-}$, blue). Conversely, the upper mode displays an opposite response, with about 50\% more spectral weight in the left-handed component than in the right-handed.

Figure~\ref{Figure2}(c) presents the imaginary part of the complex off-diagonal conductivity tensor, $\Im\{\sigma_{xy}\}$, associated with the absorptive component of the optical Hall conductivity (see Methods). The sign of $\Im\{\sigma_{xy}\}$ reflects the circular dichroism of the optical transition, distinguishing right- from left-handed contributions. As shown in Fig.~\ref{Figure2}(c), the two peaks corresponding to the electromagnons exhibit opposite signs, indicating opposite chirality, yet they possess comparable spectral weight.

To quantify the circular dichroism of the magnons, we evaluated the spectral weight ratio using a method analogous to that for a single cyclotron mode~\cite{deVisser2016}:
\begin{equation}
    A_n = 2\frac{W_n^{+} - W_n^{-}}{W_n^{+} + W_n^{-}},
\end{equation}
where $n$ refers to the lower (L) or upper (R) mode, and $W_n^{\pm}$ are the spectral weights of the right- and left-handed components, respectively. We associate the lower and upper modes with the L- and R-polarized components. The quantity $A$ reflects the circular dichroism derived from the spectral weight imbalance of each conductivity peak in the real part of $\sigma_{\pm}$.

Figure~\ref{Figure2}(d) shows the magnetic field dependence of $A_L$ (green circles) and $A_R$ (red squares). The dichroism of both modes increases approximately linearly with the magnetic field, though with slightly different slopes. The higher-energy electromagnon mode evolves more rapidly than the lower one. This trend is consistent with our transmission results, which show that the upper excitation exhibits a larger field-induced frequency shift-about 2~cm$^{-1}$ for the upper mode compared to 1~cm$^{-1}$ for the lower mode at 16~T. Notably, the dichroism exhibits a nonzero value at zero magnetic field. Typically, one would expect zero dichroism in the absence of a magnetic field, since there is no preferred direction for spin excitations. However, in NiI$_2$, the intrinsic magnetic anisotropy naturally splits the magnon modes, leading to finite dichroism even without an external field. This behavior is consistent with calculated dichroism from Raman scattering experiments~\cite{Gao2024,Song2022}.

To further probe the predicted inverse chirality of the electromagnon excitations, we performed magnetic circular dichroism (MCD) measurements in the terahertz range via chirality-resolved transmission using both left- and right-circularly polarized (LCP and RCP) light. Figures~\ref{Figure2}(e) and (f) display the field-dependent evolution of the absorption peaks for the two magnon modes under LCP and RCP excitation, respectively. Remarkably, both resonances exhibit non-reciprocal field dependence, with clear energy asymmetries in opposite field directions, confirming the chiral nature of the excitations. The 34~cm$^{-1}$ mode, associated with RCP, behaves differently compared to its LCP counterpart, and vice versa for the 37~cm$^{-1}$ mode, indicating that each magnon preferentially couples to a distinct circular polarization depending on the field direction. Furthermore, our Faraday rotation measurements show that the electromagnons exhibit asymmetric behavior with respect to $\pm 45^\circ$ polarization angles and inverted magnetic fields, consistent with the inverse chirality expected from the modified Kitaev exchange interaction model. 

To further investigate the chiral nature of the electromagnon excitations and their symmetry properties, we focus on the behavior of the circularly polarized response near zero magnetic field. Fig.~\ref{Figure2}(d–f) reveal an anomalous nonzero circular response at zero applied magnetic field, an unexpected feature for a purely antiferromagnetic system. Specifically, Fig.~\ref{Figure2}(d) demonstrates a finite circular dichroism at zero field, consistent with earlier zero-field Raman observations in NiI$_2$ that also reported opposite circular behavior for the two electromagnons~\cite{Gao2024,Song2022}. Likewise, the asymmetric evolution of the circularly polarized THz absorption around zero field in Fig.~\ref{Figure2}(e–f) further confirms the presence of an intrinsic chiral component even without external magnetic perturbation. To clarify the microscopic origin of this behavior, we performed magnetization measurements, which reveal a weak ferromagnetic moment in the helical multiferroic phase. Such a residual moment, enabled by symmetry breaking in the helical state, may originate from a small Dzyaloshinskii–Moriya–type interaction or subtle spin canting effects, which are not included in our minimal Kitaev-based Hamiltonian. While these contributions merit further investigation and incorporation into a comprehensive theoretical framework of NiI$_2$, our magneto-optical results clearly indicate that their influence is minor compared to the dominant Kitaev exchange interaction governing the chiral magnon dynamics. Taken together, these findings demonstrate that the chiral electromagnons in NiI$_2$ are primarily driven by bond-dependent Kitaev interactions, while additional small symmetry-breaking terms set the zero-field baseline of optical chirality without altering the key field-tunable behavior.


In summary, we explored the magneto-optical properties of the van der Waals multiferroic material NiI$_2$ through transmission and Faraday rotation measurements deep within the antiferromagnetic helical phase. We investigated the magnetic-field dependence of two collective magnetic excitation modes at 34~cm$^{-1}$ and 37~cm$^{-1}$, which exhibit distinct blueshifts under increasing field, along with a dichroic absolute frequency shift. We found that this behavior is isotropic with respect to the field orientation, indicating that the electromagnons in NiI$_2$ cannot be explained by a simple spin-helical model. To address this, we modeled the magnetic exchange interactions using a Kitaev Hamiltonian augmented by an $xy$-plane easy-axis term, $A_{zz}$. This model successfully accounts for the zero-field splitting of the magnon into two branches and reproduces the anomalous field dependence of the electromagnon blueshift. It also predicts inverted chirality as a function of magnetic field, which is supported by our Faraday rotation measurements that reveal opposite circular behavior of the optical conductivity for the two electromagnons. Furthermore, both the Faraday angle and MCD spectra exhibit asymmetry under magnetic field reversal, providing direct evidence for the chirality inversion expected from the Kitaev interaction.

Our results offer compelling evidence that the spin Hamiltonian of NiI$_2$ is governed primarily by Kitaev exchange interactions, in contrast to previously proposed models. Notably, the two observed electromagnons originate from a single magnon branch that is intrinsically split by symmetry-breaking mechanisms at zero magnetic field. This finding provides new insight into the interplay between the helical antiferromagnetic configuration and internal anisotropies, enriching our understanding of multiferroicity in NiI$_2$.

Our result carries broader significance in the context of topological magnetism. The Kitaev interaction, originally proposed as a route to quantum spin liquids, is now emerging as a powerful mechanism for stabilizing exotic magnetic textures, including skyrmions with higher topological charges. In contrast to conventional skyrmion systems, which rely on Dzyaloshinskii–Moriya or Ruderman–Kittel–Kasuya–Yoshida (RKKY) interactions and typically require external magnetic fields, the Kitaev interaction can drive the formation of high-order skyrmion crystals (SkX-$N$) purely via bond-dependent anisotropy. Our study supports this emerging framework by providing experimental evidence for strong Kitaev physics in NiI$_2$, complementing recent reports of a field-free SkX-2 phase in the same material~\cite{kim2024nii2}. Thus, NiI$_2$ connects the domains of quantum magnetism and topological spin textures, offering a highly tunable platform for exploring the interplay between chirality, anisotropy, and topology in low-dimensional magnets~\cite{Fert2017,Nagaosa2013,Parkin2015}. Although the present work investigated a modified Kitaev model in the helical magnetic state, the possibility of tuning the A$_{zz}$, e.g. by external strain or chemical doping, might push the system back into the intermediate skyrmion lattice state ~\cite{kim2024nii2} while keeping the multiferroicity. As such, NiI$_2$ can host a fertile ground to explore the interplay of exotic magnetic orders, such as high skyrmion lattice and ferroelectricity, namely using electrical fields to control skyrmions.

\section*{METHODS}

\subsection{Sample growth}
NiI$_2$ single crystals were grown using the chemical vapor transport (CVT) method. Nickel powder and high-purity iodine (99.99\%) were mixed in a stoichiometric ratio, with an additional 5\% iodine to ensure a complete reaction. This mixture was then sealed in a quartz tube under vacuum. The quartz tube was placed in a horizontal two-zone furnace, with the furnace ends heated to 750 $^{\circ}$C and 720 $^{\circ}$C, respectively, over a period of 6 hours. The furnace temperature was maintained for one week to promote crystal growth, followed by slow cooling to room temperature over five days. The resulting crystals were shiny grey flakes, measuring about 5 $\times$ 5 $\times$ 0.1 mm$^3$.

\subsection{Magneto-transmission}

We performed magneto-optical measurements of transmission, reflection, and Faraday angle rotation on NiI$_2$ samples. Normalized magneto-transmission measurements up to 16~T were carried out in a solenoid superconducting magnet at 4.2~K, coupled to a Bruker Vertex 80v Fourier-transform spectrometer at Laboratoire National des Champs Magnétiques Intenses (LNCMI), Grenoble, France. We probed the far-infrared regime using a Globar IR source and a liquid-He-cooled bolometer detector. The output of the spectrometer was directed through a series of light-pipe optics toward the sample, placed in a helium exchange gas environment at 4.2~K. The bolometer was located inside the sample chamber and directly behind the sample in transmission mode; in reflection mode, it was located outside the sample holder after a beam splitter. Magneto-reflection and magneto-transmission measurements were performed in the Faraday configuration, where the magnetic field is perpendicular to the sample surface and parallel to the incident light wavevector. We also conducted measurements in the Voigt configuration, where the field is parallel to the sample surface and perpendicular to the light wavevector, using a dedicated holder with two 45$^\circ$ mirrors and a sample rotator. This setup enabled rotation of the sample's \textit{ab}-plane with respect to the magnetic field to probe in-plane anisotropy. Measurements were performed at 0$^\circ$, 30$^\circ$, 60$^\circ$, and 90$^\circ$ relative to the field. The isotropic nature of the sample plane was confirmed using an incoming polarizer and an outgoing analyzer. For each magnetic field, the transmission spectrum $T(B)$ was normalized by the zero-field spectrum $T(0)$, yielding $T(B)/T(0)$. This normalization enabled accurate tracking of field-induced spectral changes.

Due to the limited dynamic range of the bolometer, a smaller iris (1~mm) was used for the reference, compared to 3~mm for the sample. Accordingly, we scaled the absolute transmission value by the square of the iris radius ratio, corresponding to a factor of 1/9. The magnon frequencies were extracted by fitting the spectra at each field using a Drude–Lorentz model in RefFit, focusing on the central frequencies $\omega_0$ of the electromagnon modes~\cite{Kuzmenko2005}.

Additional magneto-transmission measurements were performed in an optical cryostat coupled to a Bruker 70v Fourier-transform infrared spectrometer at the University of Geneva. We covered the spectral range of 15--700~cm$^{-1}$ using Globar and Hg arc-lamp sources, and gold and silicon beam splitters for the FIR and XFIR (terahertz) regimes, respectively. The sample was mounted on a custom-built optical-flow cryostat inside a superconducting split-coil magnet chamber. The magnetic field was tunable up to $\pm$7~T and aligned in the Faraday configuration. The sample temperature was controllable between 4.2~K and room temperature. The sample was affixed to the cryostat holder using silver paste or GE varnish, positioned over a 3~mm iris, and compared to a reference empty 3~mm iris. A linear translation stage allowed switching between sample and reference, enabling absolute transmission measurements, $T = I_s / I_r$, where $I_s$ and $I_r$ are the intensities from the sample and reference, respectively, measured with an IR Labs liquid-He-cooled Si bolometer. Field- and temperature-dependent spectra were recorded by measuring both sample and reference at each point.

\subsection{LSWT simulation details}

Linear spin-wave theory (LSWT) simulations were performed using the \texttt{Sunny.jl} package~\cite{Sunny2025}. It is known that the Kitaev interaction, when combined with an incommensurate magnetic ground state, cannot be treated properly using conventional LSWT due to harmonic expansions of bosonic operators~\cite{kim2024nii2, kim2023bond}. To address this, we simulated the magnon energies using the kernel polynomial method (KPM)~\cite{KPM2024} on a $1 \times 14 \times 2$ supercell. The minimal Hamiltonian of NiI$_2$ reads~\cite{kim2024nii2}:
\begin{align}
    H = &\sum_{\langle i, j \rangle_1 \in \{\alpha, \beta, \gamma \}} [J_1 \mathbf{S}_i \cdot \mathbf{S}_j + K S_i^\gamma S_j^\gamma] + \notag \\
    &\sum_{\langle i, j \rangle_n}^{3, c_1, c_2} J_n \mathbf{S}_i \cdot \mathbf{S}_j + A_{zz} \sum_i \left(S_i^z\right)^2
\end{align}
Here, $n = 1, 3$ refers to intralayer nearest neighbors, and $n = c_1, c_2$ to interlayer couplings. $K$ denotes the Kitaev interaction along $X$-, $Y$-, and $Z$-bonds (see Ref.~\cite{kim2024nii2} for bond conventions), and $A_{zz}$ is the single-ion anisotropy along the $z$-axis. We used the following parameters: $J_1 = -8.5$~meV, $K = 2.83$~meV, $J_3 = 3.13$~meV, $J_{c1} = -0.05$~meV, $J_{c2} = 1.43$~meV, and $A_{zz} = 0.32$~meV. Note that the parameter set was rescaled by 15\% from Ref.~\cite{kim2024nii2}, which is typical when simulating low-temperature spin waves from high-temperature fits.

For FTIR intensity simulations~\cite{Oshikawa2002}, we calculated:
\begin{equation}
    I(\omega) = \omega \left[S^{xx}(0, \omega) + S^{yy}(0, \omega)\right],
\end{equation}
where $S^{\alpha\beta}(Q, \omega)$ is the dynamical spin structure factor with $(\alpha, \beta) \in \{x, y, z\}$. To examine chirality, we calculated the chiral dynamical spin structure factor~\cite{Loire2011}:
\begin{equation}
    C_z(\omega) = \mathrm{Im}\left[S^{xy}(0, \omega) - S^{yx}(0, \omega)\right].
\end{equation}

\subsection{Faraday angle rotation}

Polarization-resolved measurements were performed using gold and polyethylene (PE) wire-grid polarizers for the XFIR (THz) and FIR regimes, respectively. As noted, NiI$_2$ is optically isotropic in both its normal and antiferromagnetic states. We employed the fast-protocol method from Ref.~\cite{Levallois2015,Panda2025} to measure the Faraday rotation angle. This protocol requires only four measurements, corresponding to the combinations of $\pm B$ and $P = A = \pm 45^\circ$, where $P$ and $A$ denote the polarizer and analyzer angles, respectively. In our system, the Faraday rotation angle is measured by rotating the polarizer while keeping the analyzer fixed.

The complex optical conductivity in the $ab$-plane can be written as a tensor with diagonal elements $\sigma_{xx}$ and $\sigma_{yy}$ and off-diagonal elements $\pm \sigma_{xy}$. For an isotropic response, $\sigma_{xx} = \sigma_{yy}$. The circular optical conductivities are given by $\sigma_{\pm} = \mathrm{Re}\{\sigma_{xx}\} \pm \mathrm{Im}\{\sigma_{xy}\}$, where transmission reflects $\mathrm{Re}\{\sigma_{xx}\}$ and the Faraday rotation angle is proportional to $\mathrm{Re}\{\sigma_{xy}\}$. Using the Kramers–Kronig relations and the fast-protocol method, we extract the full circular optical conductivities $\sigma_{\pm}$, and from their symmetric and antisymmetric components, we reconstruct $\sigma_{xx}$ and $\sigma_{xy}$.

\subsection{Magnetic circular dichroism}

Circularly polarized terahertz transmission was measured using a TERA K15 terahertz time-domain spectrometer (Menlo Systems, Germany) coupled to a closed-cycle SpectromagPT magneto-optic cryostat (Oxford Instruments, UK). The system operates from 1.5~K to 300~K and supports magnetic fields from $-7$~T to $+7$~T. Circularly polarized terahertz time-domain spectroscopy (THz-TDS) was performed using a quarter-wave plate to convert linearly polarized light into left- (LCP) or right-circularly polarized (RCP) light, determined by the fast axis orientation. LCP corresponds to counterclockwise rotation when the observer faces into the light source~\cite{Jackson1998}. To eliminate water vapor absorption, the entire THz path was purged with nitrogen gas inside a Plexiglas enclosure. To induce in-plane electric polarization \textbf{P}, the NiI$_2$ crystal was cooled from 80~K to 1.5~K under an in-plane field $\mathbf{H}_{\mathrm{cool}} = 4$~T~\cite{Kim2023}, which resulted in two ferroelectric domains with their opposite \textbf{P} vectors nearly perpendicular to $\mathbf{H}_{\mathrm{cool}}$. The sample (thickness 520~\textmu m) was mounted on a gold-coated copper holder with a 2.5~mm aperture and secured using Kapton tape. The chiral polarization, defined as $(I_{\mathrm{LCP}} - I_{\mathrm{RCP}})/(I_{\mathrm{LCP}} + I_{\mathrm{RCP}})$, $I$ is the intensity, is used to quantify the chiral asymmetry in absorption between LCP and RCP light.

\section*{Acknowledgements}
This research was supported by the Israel Science Foundation (ISF) through project number 666/23. N.B. and K.P. would like to acknowledge useful discussions with Dominik M Juraschek. 
The work at Sogang University was supported by G-LAMP Program of the National Research Foundation of Korea (NRF) grant funded by the Ministry of Education (No. RS-2024-00441954).
The work at Yonsei University was supported by the National Research Foundation of Korea (Grant No. NRF-2021R1A2C3004989) and the Samsung Science and Technology Foundation (Grant No. SSTF-BA2102-04).
The work at SNU was supported by the Leading Researcher Program of the National Research Foundation of Korea (Grant No. 2020R1A3B2079357) and the National Research Foundation of Korea (Grant No RS-2020-NR049405). The work at GT was supported by the US Department of Energy, Office of Science, Basic Energy Sciences, Materials Sciences and Engineering Division under award DE-SC-0018660. J.T.-G. acknowledge finantial support through the SNSF Swiss Postdoctoral Fellowships from the Swiss National Science Foundation (SNSF) under grant TMPFP2\_224378\\

\section*{Author Contributions}
K.P., F.L.M., J.D., G.L., M.O., and N.B. conducted the high-field infrared magneto transmission measurements. K.P., D.B., J.T.G., A.B.K., and N.B. performed the absolute transmission and Faraday angle measurements. K.P. and D.B. analyzed the data. Y.L. and J.G.P. prepared the single crystals. C.K., M.M., and J.G.P. performed the theoretical calculations and magnetization measurements. J.H.K., B.K., and J.H.K. carried out the MCD measurements. N.B. and J.G.P. supervised the project. K.P., N.B., C.K., and J.G.P. wrote the manuscript, with inputs from all authors. 

\section*{Competing interests}
The authors declare no competing interests.

\section*{Data availability}
The data that support the findings of this study are available from the corresponding author upon reasonable request.

\section*{Code availability}
The code that support the findings of this study are available from the corresponding author upon reasonable request.


\begin{appendices}

\section{Evolution of magnon excitations with field and easy axis model} \label{sec:appendixA}

\begin{figure*}[bth]
    \centering
    \includegraphics[width=\textwidth]{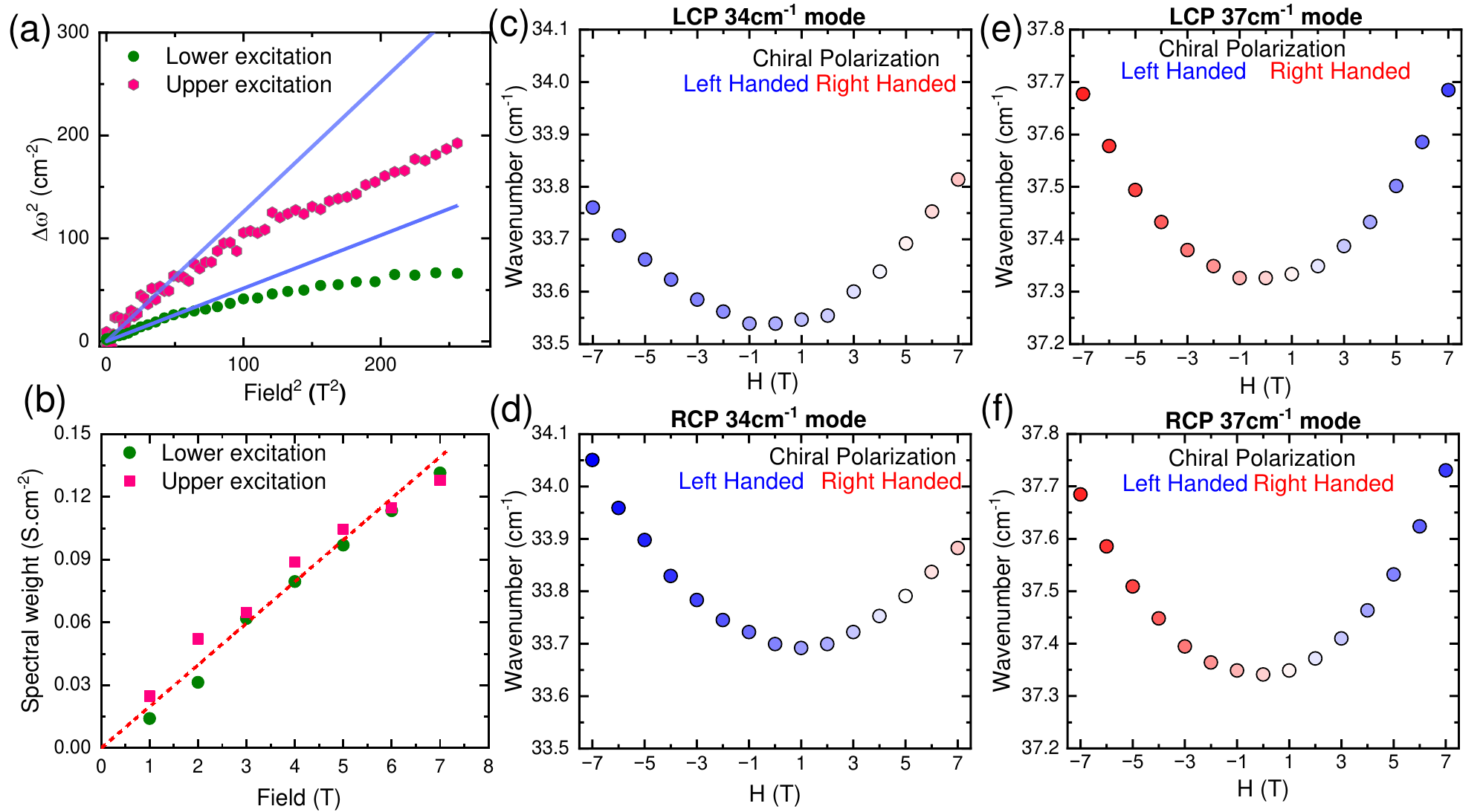}
    \captionsetup{justification=justified, singlelinecheck=false}
    \caption{
        \textbf{Supplementary Fig. S1 | Evolution of magnon central frequency and transmission in the Faraday configuration.} 
        \textbf{a,} Evolution of the magnon mode with magnetic field. The solid line represents a fit using the easy-axis model, as described in the text.  
        \textbf{b,} Spectral weight of the imaginary part of the transverse optical Hall conductivity as a function of magnetic field at $T = 4.2$~K. The solid line shows a linear fit to the data.  
        \textbf{c–f,} Peak positions of circularly polarized absorption spectra for the lower (34~cm$^{-1}$) and upper (37~cm$^{-1}$) magnon modes under left- and right-circularly polarized (LCP/RCP) terahertz radiation. The chiral polarization, defined as $(I_{\mathrm{LCP}} - I_{\mathrm{RCP}})/(I_{\mathrm{LCP}} + I_{\mathrm{RCP}})$, $I$ is the intensity.
    }
    \label{FigureS1}
\end{figure*}

Figure \ref{FigureS1}(a) presents the evolution of the square of the relative central frequency, $\Delta\omega = \omega(B)-\omega(0)$, of the two magnons as a function of the square of the applied magnetic field up to 16 T. The first electromagnon mode, centered around 34 cm$^{-1}$, shows a relatively modest blue shift, with its frequency increasing by approximately 2.5\% over the applied field range. In contrast, the second electromagnon mode, centered near 37 cm$^{-1}$, exhibits a more pronounced shift of about 7\% as the magnetic field increases to 16 T. Initial attempts to model the field dependence of the magnon modes using an easy-axis model described by the equation $\omega(B)^2 = \omega_0^2 + g\mu_B B^2$ did not accurately capture the observed frequency shifts across the full range of magnetic fields. In an easy-axis model, $\Delta\omega^2$ should be linear in $B^2$, which is not valid in our case, as can be seen in Figure~\ref{FigureS1}(a). 

\section{Spectral weight of Circular Dichroism}\label{sec:appendixB}

The field dependence of the circular dichroism can be seen also through the spectral weight of the peaks in $\Im\{\sigma_{xy}\}$. We have integrated the area below each peak in $\Im\{\sigma_{xy}\}$ and show the result of the spectral weight: 
\begin{equation}
    W_{xy} \propto \int_{\omega_{1}}^{\omega_{2}} \Im \{\sigma_{xy}(\omega)\} \, d\omega .
\end{equation}
as a function of the increasing magnetic field up to 7~T. Here we have considered $\omega_{1} = 32.34$ \cm\ and $\omega_{2} = 35.06$ \cm\ for the first peak and  $\omega_{1} = 35.06$ \cm\ and $\omega_{2} = 38.8$ \cm for the second peak. Figure~\ref{FigureS1}(b) show the spectral weight $W_{xy}$ of two excitation as a function of magnetic field where we have taken the absolute value for the area of the lower excitation. The increasing magnetic field shifts the peak position slightly, as was already observed in the transmission spectra, but also for the linear increase of $W_{xy}$ as a function of magnetic field.  

\section{THz Magnetic Circular Dichroism}\label{sec:appendixC}

Figure~\ref{FigureS1}(c)-(f) presents the peak positions of circularly polarized absorption spectra of the lower (34 cm$^{-1}$) and upper (37 cm$^{-1}$) magnon modes in NiI$_2$. Panels (c) and (d) display the magnetic field evolution of the absorption peaks for the lower mode under LCP and RCP excitation, respectively. Similarly, panels (e) and (f) show the field-dependent absorption behavior for the upper mode. The field was varied between -7 T and +7 T, and the absorption frequency shift was observed as a function of the applied magnetic field.

The absorption spectra reveal that each magnon mode exhibits distinct field-induced absorbtion depending on the polarization of the incident THz radiation. Specifically, the lower electromagnon (34 cm$^{-1}$) shows a more prominent activity under RCP, while the upper mode (37 cm$^{-1}$) exhibits a significant response to LCP. Evaluated chiral polarization reveals distinct difference between lower and upper electromagnon modes. This asymmetry highlights the chiral nature of the magnon modes and provides direct evidence for the inverse chirality predicted by the modified Kitaev interaction model.

\begin{figure*}[bth]
    \centering
    \includegraphics[width=\textwidth]{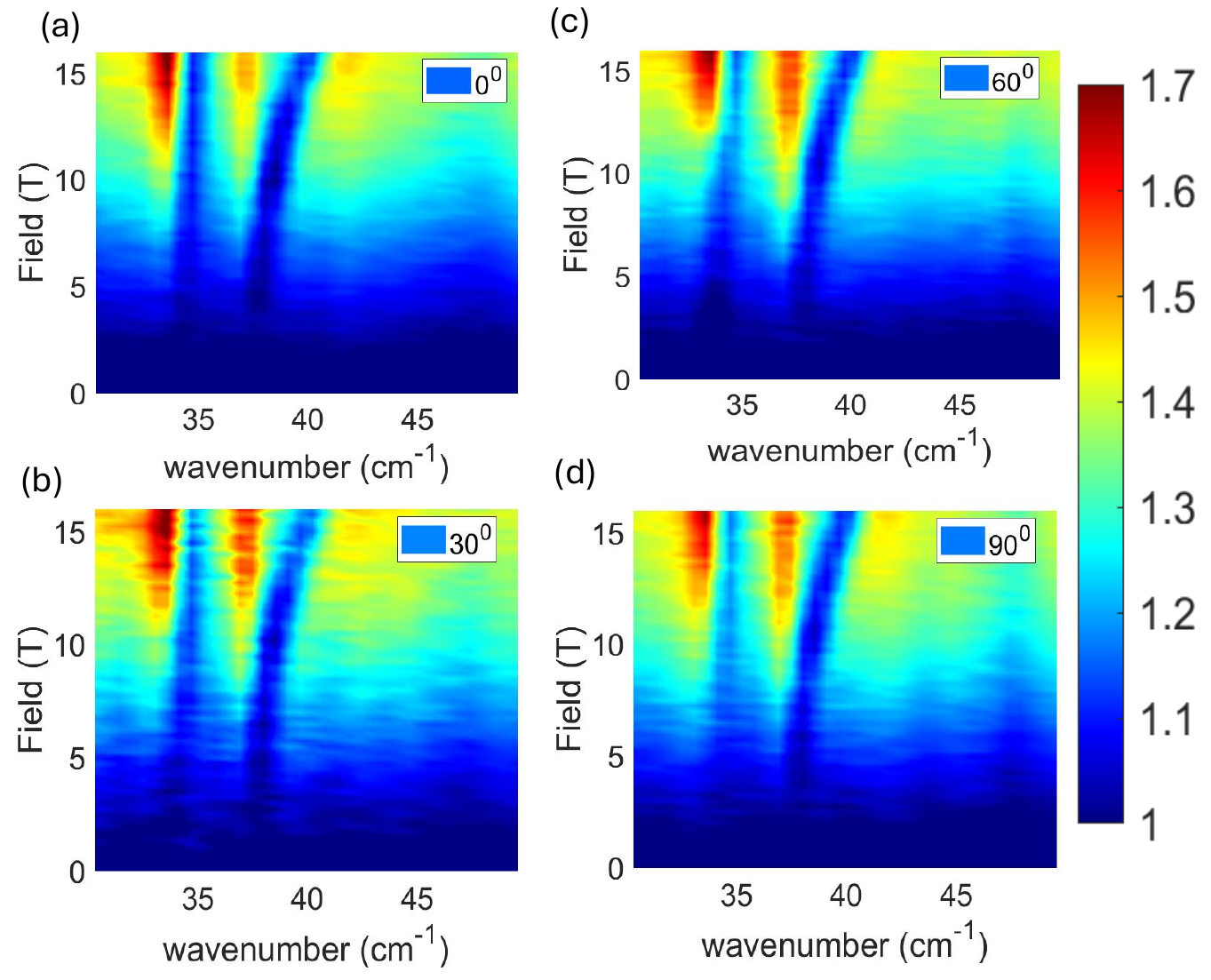}
    \captionsetup{justification=justified, singlelinecheck=false}
    \caption{
        \textbf{Supplementary Fig. S2 | Transmission in the Voigt configuration.}    
        \textbf{a–d,} Magneto-optical response of NiI$_2$ in the far-infrared spectral range at $T = 4.2$~K in the Voigt geometry ($B \perp E$).
    }
    \label{FigureS2}
\end{figure*}

\section{Angular dependence of magnon excitations}\label{sec:appendixD}

Figure~\ref{FigureS2}(c-f) shows a 2D color map of the field-dependent transmission spectra \(T(B)\), measured in the Voigt configuration for the 4 different a-b plane orientations compared to the applied magnetic field and normalized to the zero-field transmission \(T(0)\). By comparing the different Voigt configuration spectra and that of the Faraday configuration, we see no significant variation in the magnons' central frequency with the magnetic field. These findings suggest that the magnon excitation is isotropic for the investigated directions or that we have not measured the magnons in their respective easy axis. This observation is consistent with the magnetic structure in the multiferroic phase, characterized by a helical modulation with an incommensurate wave vector \(\mathbf{q} \approx (0.138, 0, 1.457)\). The modulation vector forms an angle of approximately \(71^\circ\) with the \(ab\)-plane, and the spin plane is inclined at about \(55^\circ\) relative to the \(c\)-axis~\cite{Kuindersma1981}.

\begin{figure*}[bth]
    \centering
    \includegraphics[width=\textwidth]{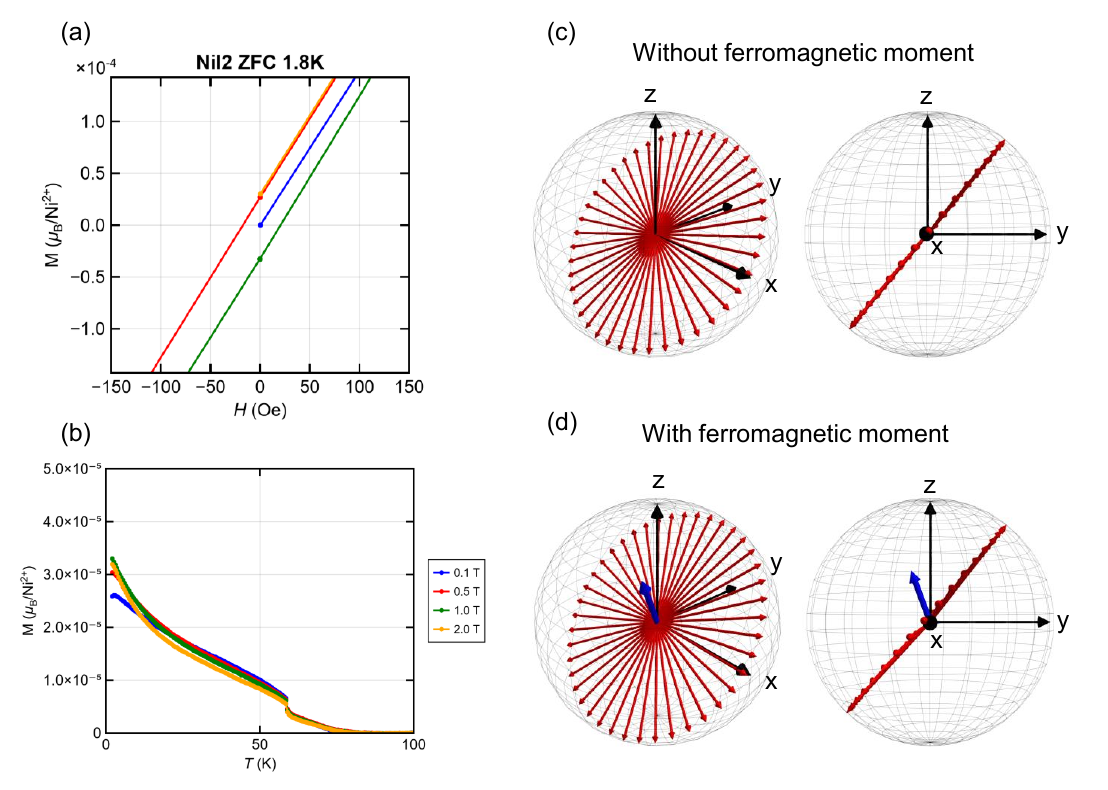}
    \captionsetup{justification=justified, singlelinecheck=false}
     \caption{
        \textbf{Fig. 4 | Hidden ferromagnetic moments in NiI$_2$.} 
        \textbf{a,} Field-dependent magnetization measurements of NiI$_2$ at $T = 1.8$~K with $H \parallel c$. 
        \textbf{b,} Temperature dependence of magnetization in NiI$_2$, measured after field cooling under a finite magnetic field with $H \parallel c$. Each color represents a distinct magnetic field. 
        \textbf{c, d,} Bloch sphere representation of the magnetic moments in NiI$_2$. \textbf{c} shows the ideal magnetic structure of NiI$_2$ without ferromagnetic moments, while \textbf{d} depicts the helical magnetic order with weak ferromagnetic moments. The blue arrow indicates the direction of the weak ferromagnetic moments.
    }
    \label{FigureS3}
\end{figure*}

\subsection{Emergence of Weak Ferromagnetism and Zero-Field Chirality of Electromagnons}\label{sec:appendixE}
However, one intriguing feature that LSWT could not explain is the finite chirality of the electromagnons in the absence of an external magnetic field. The presence of chirality at zero field implies spontaneous time-reversal symmetry breaking. In noncollinear magnetic systems, such symmetry breaking is often associated with the emergence of a weak ferromagnetic moment. To examine this possibility, we carefully measured the magnetization and identified a tiny ferromagnetic moment of ~10$^{-5}$ $\mu_{B}/f.u.$, which has not been reported previously (see Fig.~\ref{FigureS3}). The temperature dependence shows that this weak ferromagnetic moment persists up to 75 K, indicating its close connection to the underlying magnetic order. Moreover, field-dependent magnetization measurements reveal a clear hysteresis loop. Taken together, these observations strongly suggest that the weak ferromagnetic moment originates from a subtle Dzyaloshinskii-Moriya interaction enabled by the multiferroic phase transition. Crucially, this hidden ferromagnetism provides the microscopic origin of the observed zero-field chirality of the electromagnons, directly linking the broken time reversal symmetry to the chiral magnon dynamics.

\end{appendices}


\bibliography{ref_Nature}

\end{document}